\documentclass{elsart}

\usepackage{graphicx}
\usepackage[dvips]{thumbpdf}
\usepackage{amssymb}

\begin{document}

\begin{frontmatter}

\title{A relativistic framework to determine the nuclear transparency
from $A(p,2p)$ reactions}

\author{B. Van Overmeire}\ead{Bart.VanOvermeire@UGent.be},
\author{J. Ryckebusch}

\address{Department of Subatomic and Radiation Physics,
Ghent University, Proeftuinstraat 86, B-9000 Gent, Belgium}

\begin{abstract}
A relativistic framework for computing the nuclear transparency
extracted from $A(p,2p)$ scattering processes is presented.  The model
accounts for the initial- and final-state interactions (IFSI) within the
relativistic multiple-scattering Glauber approximation (RMSGA).  For the
description of color transparency, two existing models are used.  The
nuclear filtering mechanism is implemented as a possible explanation for
the oscillatory energy dependence of the transparency.  Results are
presented for the target nuclei \nuc{7}{Li}, \nuc{12}{C}, \nuc{27}{Al},
and \nuc{63}{Cu}.  An approximated, computationally less intensive
version of the RMSGA framework is found to be sufficiently accurate for
the calculation of the nuclear transparency.  After including the
nuclear filtering and color transparency mechanisms, our calculations
are in acceptable agreement with the data.
\end{abstract}

\begin{keyword}
$A(p,2p)$ transparency \sep Glauber theory \sep color transparency \sep 
nuclear filtering
\PACS 13.75.Cs \sep 13.85.Dz \sep 25.40.Ve \sep 11.80.La
\end{keyword}
\end{frontmatter}

\section{Introduction}
\label{sec:intro}

The transition region between nucleon-meson (hadronic) and quark-gluon
(partonic) degrees of freedom is a topic of longstanding interest in
nuclear physics.  A promising observable to map this transition is the
transparency of the nuclear medium to the propagation of hadrons.  In
$A(p,2p)$ experiments, the nuclear transparency is defined as the ratio
of the cross section per nucleon to the hydrogen one.  Accordingly, the
nuclear transparency is a measure for the attenuation effects of the
spectator nucleons on the impinging and outgoing protons.

In the conventional Glauber picture \cite{glauber59}, the nuclear
transparency extracted from $A(p,2p)$ reactions is predicted to be
rather constant for incoming momentum larger than a few GeV/c.  The
color transparency (CT) phenomenon suggests an anomalously large
transmission probability of protons through nuclei
\cite{mueller82,brodsky82} and leads to a nuclear transparency that
increases with incoming momentum.  The experimental $A(p,2p)$ data
\cite{carroll88,mardor98,leksanov01} suggest a CT-like increase in the
transparency for impinging proton momenta between $5$ and $10$~GeV/c.
For higher momenta the nuclear transparency falls back to the Glauber
level.  This oscillatory energy dependence is not unique to the
$A(p,2p)$ nuclear transparency: it has been observed or hinted at in
$pp$ elastic scattering \cite{hendry74}, elastic $\pi p$ fixed-angle
scattering \cite{pi-pscatt,jain02}, pion photoproduction
\cite{jain02,zhu0305}, and deuteron photodisintegration
\cite{deut_photo_exp,frankfurt00}. 

One possible interpretation of this energy dependence of the $A(p,2p)$
transparency is provided by the presence of two terms in the free $pp$
scattering amplitude \cite{ralston88,brodsky88}.  Ralston and Pire
\cite{ralston88} suggested a combination of the following two
components.  First, the quark-counting component which follows the
dimensional scaling law \cite{countingrule} and represents a small
object (a point-like configuration, PLC).  Second, the Landshoff
component \cite{landshoff74} which is associated with normal-sized
configurations.  The interference between these two amplitudes induces
the oscillation of the $pp$ cross section about the scaling behavior.
Inside the nuclear medium, the Landshoff component will be suppressed
because of the strong interactions with the spectator nucleons, while
the quark-counting component will escape the nucleus with relatively
small attenuation due to CT.  This phenomenon is called ``nuclear
filtering'' (NF) \cite{jain96}: the nucleus filters away the
normal-sized components in the hadron wave functions.  Accordingly, the
nucleus plays an active role in selecting small-sized components.  The
NF model is able to reproduce qualitatively the observed bump in the
$A(p,2p)$ transparency.

Recently, a relativistic and cross-section factorized framework for
$A(p,2p)$ reactions has been proposed \cite{vanovermeire06}.  In this
Letter, this formalism is extended to incorporate the Ralston-Pire model
for the $pp$ scattering amplitude and the concepts of CT and NF.  For
the description of the initial- and final-state interactions (IFSI) of
the impinging and two outgoing protons, we adopt the relativistic
multiple-scattering Glauber approximation (RMSGA) \cite{vanovermeire06}.
This relativistic extension of the Glauber model, which was originally
developed to describe $A(e,e'p)$ observables \cite{ryckebusch03,lava04},
models the IFSI as consecutive cumulative scatterings with the
individual spectator nucleons in the nucleus.  To compute the effects of
CT, we consider the quantum diffusion model of Ref.~\cite{farrar88} and
an alternative treatment presented in Ref.~\cite{jennings91}.  The
comparison between the two CT models is made in a consistent way.

This Letter is organized as follows.  In Section~\ref{sec:formalism},
the RMSGA $A(p,2p)$ formalism of Ref.~\cite{vanovermeire06} is extended
to take into account both the quark-counting and the Landshoff
contribution to the hard $pp$ scattering amplitude.  A numerically
convenient approximate form of the RMSGA framework is introduced.
Further, we describe two different methods to deal with CT, thereby
indicating the similarities and differences.  Section~\ref{sec:results}
presents the nuclear transparency results for the target nuclei
\nuc{7}{Li}, \nuc{12}{C}, \nuc{27}{Al}, and \nuc{63}{Cu}.  The effect of
IFSI, CT and NF is discussed.  Furthermore, the accuracy of the
approximated RMSGA approach is investigated and the two CT treatments
are compared.  Finally, Section~\ref{sec:concl} summarizes our findings
and states our conclusions.

\section{Formalism}
\label{sec:formalism}

In the Ralston-Pire approach \cite{ralston82}, the spin-averaged $pp$
scattering amplitude consists of the quark-counting (QC) and the
Landshoff (L) contribution:
\begin{equation}
{\mathcal M}^{pp} = {\mathcal M}^{pp}_{\mathrm{QC}} +
{\mathcal M}^{pp}_{\mathrm{L}} 
\; .
\label{eq:RP_pp_matr_elt}
\end{equation}
The ${\mathcal M}^{pp}_{\mathrm{QC}}$ term results in a differential
cross section that scales like $d\sigma^{pp}/dt \propto s^{-10}$ and the
Landshoff term can be related to it through
\begin{equation}
{\mathcal M}^{pp}_{\mathrm{L}} =
\frac{\rho_{1}}{2} \:
\sqrt{\frac{s}{1~GeV^2}} \:
\e^{\pm i \bigl( \phi \left( s \right) + \delta_{1} \bigr)} \:
{\mathcal M}^{pp}_{\mathrm{QC}} \; .
\label{eq:L_pp_ampl}
\end{equation}
Here, $s$ and $t$ are the Mandelstam variables.  Further, $\rho_{1} =
0.08$, $\phi \left( s \right) = \frac{\pi}{0.06} \, \ln \left\{ \ln
\left[ \frac{s}{0.01 \, GeV^{2}} \right] \right\}$, and $\delta_{1} = -
2.0$.  These values were determined from a fit to the $pp$ data at
$90^{\circ}$ \cite{ralston82}.  The sole parameter which remains
undetermined is the sign of the phase difference $\phi \left( s \right)
+ \delta_{1}$ between the quark-counting and the Landshoff term.
Therefore, both signs will be used in the calculations.

Incorporating the Ralston-Pire approach and the NF mechanism into the
$A(p,2p)$ formalism of Ref.~\cite{vanovermeire06}, the amplitude for the
$p \left( E_{p1}, \vec{p}_{1}, m_{s1i} \right) + A \left( E_{A},
\vec{k}_{A}, 0^+ \right) \longrightarrow p \left( E_{k1}, \vec{k}_{1},
m_{s1f} \right) + p \left( E_{k2}, \vec{k}_{2}, m_{s2f} \right) +
\mbox{A-1} \left( E_{A-1}, \vec{k}_{A-1}, J_R M_R \right)$ reaction
becomes
\begin{eqnarray}
{\mathcal M}_{fi}^{(p,2p)} & = & \sum_{m_{s}}
\left( {\mathcal M}^{pp}_{\mathrm{QC}}
\right)_{m_{s1i}, m_{s}, m_{s1f}, m_{s2f}} \:
\bar{u} ( \vec{p}_{m}, m_{s} )
\phi_{\alpha_{1}}^{\mathrm{RMSGA+CT}} \left( \vec{p}_{m} \right)
\nonumber \\ & & + \sum_{m_{s}}
\left( {\mathcal M}^{pp}_{\mathrm{L}}
\right)_{m_{s1i}, m_{s}, m_{s1f}, m_{s2f}} \:
\bar{u} ( \vec{p}_{m}, m_{s} )
\phi_{\alpha_{1}}^{\mathrm{RMSGA}} \left( \vec{p}_{m} \right)
\; ,
\label{eq:RP_p2p_matr_elt}
\end{eqnarray}
where $\vec{p}_{m} = \vec{k}_{1} + \vec{k}_{2} - \vec{p}_{1}$ is the
missing momentum and $\alpha_{1}$ refers to the state wherein the struck
proton resided.  In this expression, the effect of IFSI is accounted for
through the distorted momentum-space wave functions 
$\phi_{\alpha_{1}}^{\mathrm{RMSGA+CT}} \left( \vec{p}_{m} \right)$ and
$\phi_{\alpha_{1}}^{\mathrm{RMSGA}} \left( \vec{p}_{m} \right)$.  Since
the quark-counting term is associated with PLCs, the corresponding
momentum-space wave function $\phi_{\alpha_{1}}^{\mathrm{RMSGA+CT}}
\left( \vec{p}_{m} \right)$ includes the effect of CT.  The Landshoff
term, on the other hand, corresponds with a hadron of normal size.
Consequently, the IFSI can be computed in standard Glauber theory.
Using the spin-averaged $pp$ matrix element of
Eq.~(\ref{eq:RP_pp_matr_elt}), the squared $A(p,2p)$ matrix element for
knockout from the $\alpha_{1}$ shell can be cast in the form
\begin{eqnarray}
\overline{\sum_{if}} \left| {\mathcal M}_{fi}^{(p,2p)} \right|^{2}
= \sum_{m,m_{s}} \Biggl\{
\left| {\mathcal M}^{pp}_{\mathrm{QC}} \right|^{2}
\left|
\bar{u} ( \vec{p}_{m}, m_{s} ) \:
\phi_{\alpha_1}^{\mathrm{RMSGA+CT}} \left( \vec{p}_{m} \right)
\right|^{2}
\nonumber \\ +
2 Re \biggl[ 
{\mathcal M}^{pp}_{\mathrm{QC}}
\left( {\mathcal M}^{pp}_{\mathrm{L}} \right)^{*}
\bar{u} ( \vec{p}_{m}, m_{s} ) \:
\phi_{\alpha_1}^{\mathrm{RMSGA+CT}} \left( \vec{p}_{m} \right)
\nonumber \\ \times
\left(
\bar{u} ( \vec{p}_{m}, m_{s} ) \:
\phi_{\alpha_1}^{\mathrm{RMSGA}} \left( \vec{p}_{m} \right)
\right)^{*}
\biggr]
\nonumber \\ +
\left| {\mathcal M}^{pp}_{\mathrm{L}} \right|^{2}
\left|
\bar{u} ( \vec{p}_{m}, m_{s} ) \:
\phi_{\alpha_1}^{\mathrm{RMSGA}} \left( \vec{p}_{m} \right)
\right|^{2}
\Biggr\}
\; ,
\label{eq:RP_p2p_squared_matr_elt}
\end{eqnarray}
with $m$ the struck nucleon's generalized angular momentum quantum
number.  The differential cross section is obtained as an incoherent sum
of the squared matrix elements over all proton levels $\alpha_{1}$,
thereby factoring in the occupation number of every level.  The
momentum-space wave function is defined as \cite{vanovermeire06}
\begin{equation}
\phi_{\alpha_1}^{\mathrm{RMSGA(+CT)}} \left( \vec{p}_{m} \right)
= \int d \vec{r}
\e^{- i \vec{p}_{m} \cdot \vec{r}}
\phi_{\alpha_1} \left( \vec{r} \right)
\widehat{\mathcal{S}}_{\mathrm{IFSI}}^{\mathrm{RMSGA(+CT)}}
\left( \vec{r} \right)
\; ,
\label{eq:dist_mom_space_wave_fctn}
\end{equation}
where the relativistic bound-state wave function $\phi_{\alpha_1} \left(
\vec{r} \right)$ is computed in the Hartree approximation to the $\sigma
- \omega$ model \cite{serot86}, using the W1 parametrization for the
different field strengths \cite{furnstahl97}.  The
$\widehat{\mathcal{S}}_{\mathrm{IFSI}}^{\mathrm{RMSGA(+CT)}} \left(
\vec{r} \right)$ operator accounts for the IFSI effects and is the
subject of the forthcoming discussion.  Hereafter, results obtained on
the basis of Eq.~(\ref{eq:RP_p2p_squared_matr_elt}) are dubbed
RMSGA+CT+NF. 

In the RMSGA framework, the IFSI operator takes on the form
\cite{vanovermeire06}
\begin{eqnarray}
\widehat{\mathcal{S}}_{\mathrm{IFSI}}^{\mathrm{RMSGA}}
\left( \vec{r} \right) = 
\prod_{j=2}^{A}
\Biggl\{ & &
\int d \vec{r}_j
\left| \phi_{\alpha_j} \left( \vec{r}_j \right) \right|^2 
\left[ 1 - \Gamma_{pN} \left(p_{1}, \vec{b} - \vec{b_j} \right)
\theta \left( z - z_j \right) \right]
\nonumber \\ & & \times
\left[ 1 - \Gamma_{pN} \left(k_{1}, \vec{b} \, ' - \vec{b_j} \, '
\right)
\theta \left( z_j \, ' - z \, ' \right) \right] 
\nonumber \\ & & \times
\left[ 1 - \Gamma_{pN} \left(k_{2}, \vec{b} \, '' - \vec{b_j} \, ''
\right)
\theta \left( z_j \, '' - z \, '' \right) \right]
\Biggr\}
\; ,
\label{eq:RMSGA_IFSI_factor}
\end{eqnarray}
where $\vec{r}$ denotes the point of collision between the struck and
incoming proton, and $\vec{r}_j$ ($j = 2, \ldots, A$) are the positions
of the frozen spectator nucleons.  Further, the $z$ axis lies along
$\vec{p}_{1}$, $z \, '$ along $\vec{k}_{1}$, and $z \, ''$ along
$\vec{k}_{2}$.  The $\vec{b}$, $\vec{b} \, '$, and $\vec{b} \, ''$
planes are perpendicular to these proton momenta.  Reflecting the
diffractive nature of $pN$ collisions at GeV energies, the profile
function for $pN$ scattering is parametrized as
\begin{equation}
\Gamma_{pN} \left( k, \vec{b} \right)
= \frac{{\sigma^{\mathrm{tot}}_{pN} \left( k \right)}
\left(1 - i {\epsilon_{pN}} \left( k \right) \right)}
{4\pi \left( \beta_{pN} \left( k \right) \right)^{2}} \:
\mathrm{exp} \left( - \frac{\vec{b}^{2}}
{2 \left( \beta_{pN} \left( k \right) \right)^{2}}  \right)
\; .
\label{eq:profile_fctn}
\end{equation}
The total $pN$ cross section $\sigma^{\mathrm{tot}}_{pN} \left( k
\right)$, the slope parameter $\beta_{pN} \left( k \right)$, and the
ratio of the real to the imaginary part of the scattering amplitude
$\epsilon_{pN} \left( k \right)$ depend on the proton momentum $k$.  In
the numerical calculations, their values are obtained through
interpolation of the $NN$ scattering database from the Particle Data
Group \cite{pdg}.

Eq.~(\ref{eq:RMSGA_IFSI_factor}) is a genuine $A$-body operator and the
integration over the coordinates of the spectator nucleons makes its
numerical evaluation very challenging.  The IFSI operator can be
rewritten in a numerically more convenient form by adopting the
following approximations.  First, the squared wave functions of the
spectator protons (neutrons) are replaced by $1 / \left( Z - 1 \right)$
($1 / N$) times the proton (neutron) density of the residual nucleus.
Second, one assumes that these densities are slowly varying functions of
$\vec{b}$, while $\Gamma_{pN} \left( k, \vec{b} \right)$ is sharply
peaked at $\vec{b} = \vec{0}$.  This allows one to approximate the IFSI
operator of Eq.~(\ref{eq:RMSGA_IFSI_factor}) by the one-body operator
\cite{joachain}:
\begin{eqnarray}
\widehat{\mathcal{S}}_{\mathrm{IFSI}}^{\mathrm{RMSGA^{\prime}}}
\left( \vec{r} \right) = \prod_{N=p,n} & &
\e^{- \frac{1}{2} \sigma^{\mathrm{tot}}_{pN} \left( p_{1} \right)
(1 - i {\epsilon_{pN} \left( p_{1} \right)}) 
\: \int_{- \infty}^{z} dz_j \,
\rho_{N} (\vec{b}, z_j)}
\:
\nonumber \\ & & \times
\e^{- \frac{1}{2} \sigma^{\mathrm{tot}}_{pN} \left( k_{1} \right)
(1 - i {\epsilon_{pN} \left( k_{1} \right)}) 
\: \int_{z \, '}^{+ \infty} dz_j \, ' \,
\rho_{N} (\vec{b} \, ', z_j \, ')}
\:
\nonumber \\ & & \times
\e^{- \frac{1}{2} \sigma^{\mathrm{tot}}_{pN} \left( k_{2} \right)
(1 - i {\epsilon_{pN} \left( k_{2} \right)}) 
\: \int_{z \, ''}^{+ \infty} dz_j \, '' \,
\rho_{N} (\vec{b} \, '', z_j \, '')}
\; .
\label{eq:REA_IFSI_factor}
\end{eqnarray}
Here, $\rho_{p}$ and $\rho_{n}$ are the proton and neutron density of
the residual nucleus.  They reflect the spatial distribution of the
scattering centers inside this nucleus.  Henceforth, calculations based
on Eq.~(\ref{eq:REA_IFSI_factor}) are labeled as RMSGA$^{\prime}$.

The essential assumption of CT is that the impinging proton compresses
to a PLC as it hits a target nucleon, after which the outgoing protons
expand from PLCs to normal-sized objects as they move through the
nucleus.  To account for the reduced interaction of a PLC with the
nuclear medium, the total cross sections $\sigma^{\mathrm{tot}}_{pN}$ in
Eqs.~(\ref{eq:profile_fctn}) and (\ref{eq:REA_IFSI_factor}) are replaced
by effective ones.  In the partonic model of Farrar et al.
\cite{farrar88} (denoted by FLFS), the interaction cross section is
argued to be
\begin{equation}
\sigma_{pN}^{\mathrm{FLFS}} (p, Z) = \sigma^{\mathrm{tot}}_{pN}
\:
\bigg\{
\Big[
\frac{Z}{l_h} +
\frac{\langle n^2 k^2_t \rangle}{|t|} \Big( 1 - \frac{Z}{l_h} \Big)
\Big] \theta (l_h - Z)
+ \theta (Z - l_h)
\bigg\}
\; .
\label{eq:FLFS_eff_sigma}
\end{equation}
Here, $Z$ is the distance from the hard interaction point along the
trajectory of the particle, $n = 3$ is the number of constituents in the
proton, and $\langle k^2_t \rangle^{1/2} = 0.35$~GeV/c is the average
transverse momentum of a parton in a hadron.  The quantity $l_h \simeq 2
p / \Delta M^2$ is the hadronic expansion length, i.e., the propagation
distance at which an expanding hadron reaches its normal hadronic size,
and depends on the hadron momentum $p$ and the squared mass difference
$\Delta M^2$ between the intermediate PLC and the normal-sized hadron.
It is commonly assumed that $0.7 \leq \Delta M^2 \leq
1.1$~(GeV/c$^2$)$^2$ are reasonable values.

Starting from a hadronic picture, Jennings and Miller \cite{jennings91}
suggested the following alternative expression for the effective cross
section
\begin{equation}
\sigma_{pN}^{\mathrm{JM}} (p, Z) = \sigma^{\mathrm{tot}}_{pN}
\:
\Big(
1 - \frac{p}{p^{*}} \e^{i (p - p^{*}) Z}
\Big)
\; ,
\label{eq:JM_eff_sigma}
\end{equation}
with $p$ the proton momentum and $p^{*}$ the momentum of a baryon
resonance with a complex mass $M^{*}$ and the same energy as the
nucleon, i.e., $(p^{*})^2 = p^2 + M_p^2 - (M^{*})^2$.  The expression
for the effective cross section emanates from the intermediate PLC being
a superposition of the nucleon ground state and a nucleon resonance.
The imaginary part of $M^{*}$ ensures the decay of the intermediate
state to an asymptotically free, normal-sized proton.  Like the FLFS
approach of Eq.~(\ref{eq:FLFS_eff_sigma}), Eq.~(\ref{eq:JM_eff_sigma})
considers one excited state in the PLC.  It is worth noting that both
the FLFS and JM model take into account the suppression of interaction
in the collision point and the time evolution of the PLC to a
normal-sized proton during its propagation through the nucleus.

In our numerical calculations, we will also consider the standard
RMSGA+CT picture.  In this scenario, the entire wave packet of the
incoming and outgoing protons is assumed to be in a PLC, which
propagates through a passive nuclear medium.  This amounts to neglecting
the Landshoff term in the amplitudes ${\mathcal M}^{pp}$ and ${\mathcal
M}_{fi}^{(p,2p)}$ of Eqs.~(\ref{eq:RP_pp_matr_elt}) and
(\ref{eq:RP_p2p_matr_elt}).  Finally, in the standard RMSGA
calculations, both the Landshoff term and CT effects are neglected.

For the free $pp$ scattering cross section $d\sigma^{pp}/dt$, the
parametrization as presented in Ref.~\cite{yaron02} is used.  This
parametrization combines the $\theta_{c.m.}$ dependence suggested by
\cite{sivers76} and the Ralston-Pire separation of
Eq.~(\ref{eq:L_pp_ampl}).  

\section{Nuclear transparency results}
\label{sec:results}

The nuclear transparency is computed as the ratio of the cross sections
including and excluding IFSI effects:
\begin{equation}
T = \frac{\sigma^{(p,2p)}}{\sigma^{(p,2p)}_{\mathrm{RPWIA}}}
\; .
\label{eq:T_def}
\end{equation}
The relativistic plane wave approximation (RPWIA) limit is reached by
setting the IFSI operator $\widehat{\mathcal{S}}_{\mathrm{IFSI}} \left(
\vec{r} \right)$ equal to one in
Eq.~(\ref{eq:dist_mom_space_wave_fctn}), and corresponds with a
calculation which ignores IFSI.  The numerator and denominator of
Eq.~(\ref{eq:T_def}) are obtained by integrating the corresponding
differential cross sections over the phase space defined by the
kinematic cuts.  In our calculations, we adopted identical cuts as in
the experiments \cite{carroll88,mardor98,leksanov01} and assumed a flat
experimental acceptance within the kinematical ranges for each data
point.

We consider the experimental nuclear transparency values as presented in
\cite{aclander04}.  The incident lab momentum varies from $5.9$ to
$14.4$~GeV/c and the scattering angle is near $90^{\circ}$ in the $pp$
center of mass.  The Mandelstam variable $| t | \simeq \left( s - 4
M_p^2 \right) / 2$ extends from $4.7$ to $12.7$~(GeV/c)$^2$.  In
Ref.~\cite{aclander04}, the originally published values of Carroll et
al. \cite{carroll88} were rescaled using the improved nuclear momentum
distributions of Ref.~\cite{degliAtti96}, thereby making them consistent
with the data of Refs.~\cite{mardor98,leksanov01}.

\begin{figure*}[p]
\begin{center}
\includegraphics[width=0.75\textwidth]{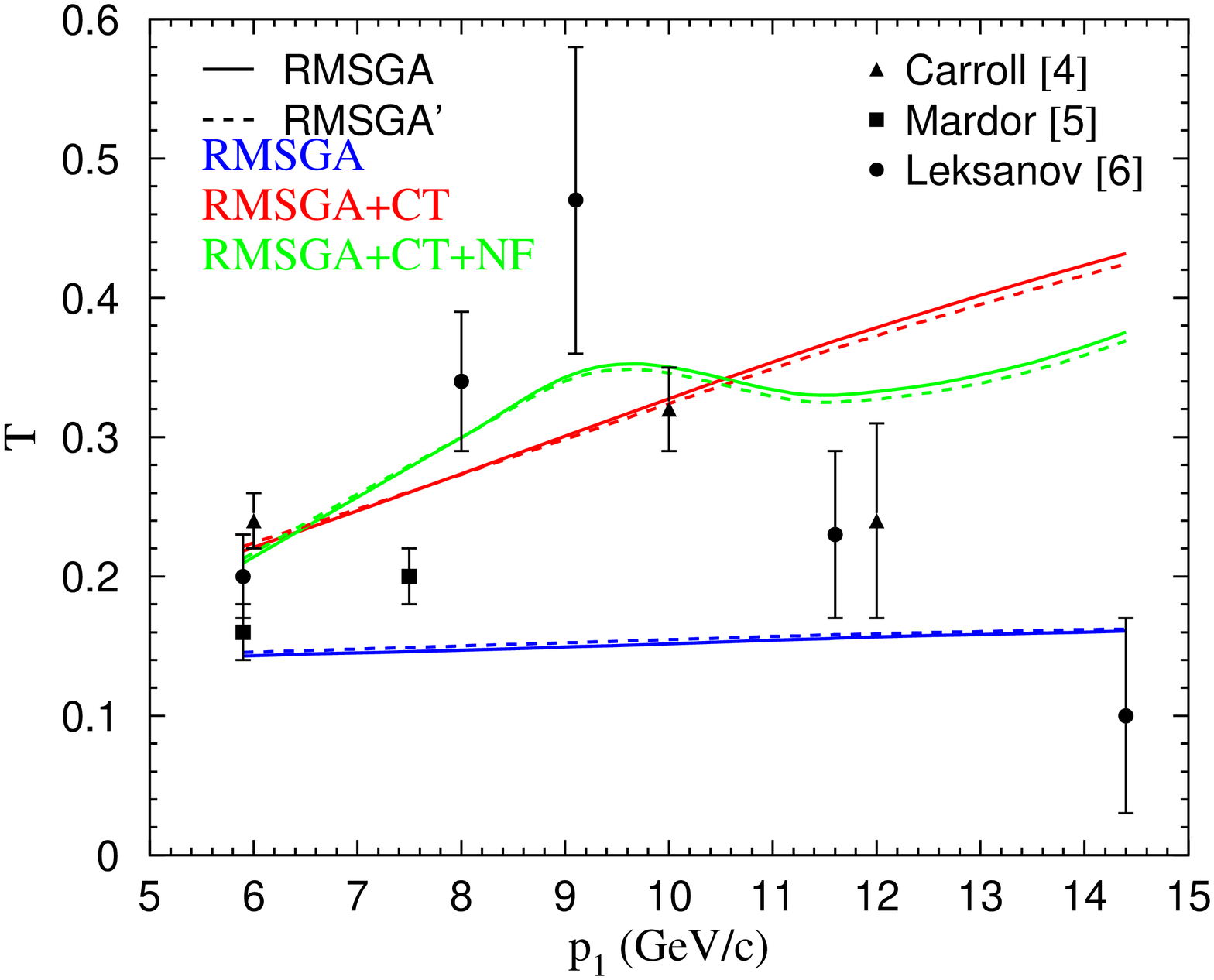}
\caption{The nuclear transparency for the \nuc{12}{C}$(p,2p)$ reaction
  as a function of the incoming lab momentum $p_1$.  The full RMSGA
  (solid lines) are compared to the RMSGA$^{\prime}$ (dashed lines)
  results.  The different curves represent the RMSGA, RMSGA+CT and
  RMSGA+CT+NF calculations.  The CT effects are calculated in the FLFS
  model \cite{farrar88} with $\Delta M^2 = 0.7$~(GeV/c$^2$)$^2$ and the
  results including the mechanism of NF are obtained using the positive
  sign of $\phi \left( s \right) + \delta_{1}$.  Data are from
  Refs.~\cite{carroll88,mardor98,leksanov01}.}
\label{fig:trans.12c.RMSGAvsREA}
\end{center}
\end{figure*}

First, we address the energy dependence of the \nuc{12}{C}$(p,2p)$
transparency and study the role of IFSI, CT and NF.  Further, we use the
\nuc{12}{C}$(p,2p)$ calculations as a test case to determine the
accuracy of the IFSI operator of Eq.~(\ref{eq:REA_IFSI_factor}) relative
to the expression of Eq.~(\ref{eq:RMSGA_IFSI_factor}).
Fig.~\ref{fig:trans.12c.RMSGAvsREA} displays the \nuc{12}{C}
transparency as a function of the incoming proton momentum $p_{1}$.  The
solid curves represent the full RMSGA calculations, whereas the
RMSGA$^{\prime}$ results are shown as dashed curves.  Three different
scenarios were considered.  As expected, the standard RMSGA calculations
lead to a nuclear transparency that is almost independent of the beam
momentum.  The main effect of the IFSI is to reduce the nuclear
transparency from the asymptotic value of $1$ to $\sim 0.15$.  The
inclusion of CT effects produces a transparency linearly rising with
energy.  The increase relative to the RMSGA result is highly dependent
on the adopted model and corresponding parameters for CT.  The curves
including CT shown in Fig.~\ref{fig:trans.12c.RMSGAvsREA} adopt the FLFS
model with $\Delta M^2 = 0.7$~(GeV/c$^2$)$^2$.  The increase of the
transparency is consistent with the data in the range $5$--$10$~GeV/c,
but the RMSGA+CT picture fails to explain the drop in the transparency
at higher momenta.  Our RMSGA and RMSGA+CT predictions confirm the
results of \cite{lee92}.  A better agreement with the data is obtained
when adding the mechanism of NF.  Compared to the RMSGA+CT results, the
transparency is increased at intermediate momenta ($5$--$10$~GeV/c) and
decreased at higher momenta, two effects which improve the description
of the data.  A similar result was obtained in Ref.~\cite{jennings92b}
where the JM model of CT was used.  

Concerning the comparison of the RMSGA and RMSGA$^{\prime}$ results, it
can be inferred from Fig.~\ref{fig:trans.12c.RMSGAvsREA} that both
approaches yield nearly identical results which differ at the $2$--$3\%$
level.  Consequently, the operator of Eq.~(\ref{eq:REA_IFSI_factor}) is
considered sufficiently accurate for the calculation of the nuclear
transparency and will be used in the remainder of this work.  We wish to
stress that the computational cost of Eq.~(\ref{eq:REA_IFSI_factor}) is 
about a factor of $10^3$ lower than the full-blown RMSGA operator of
Eq.~(\ref{eq:RMSGA_IFSI_factor}).

\begin{figure*}[p]
\begin{center}
\includegraphics[width=0.75\textwidth]{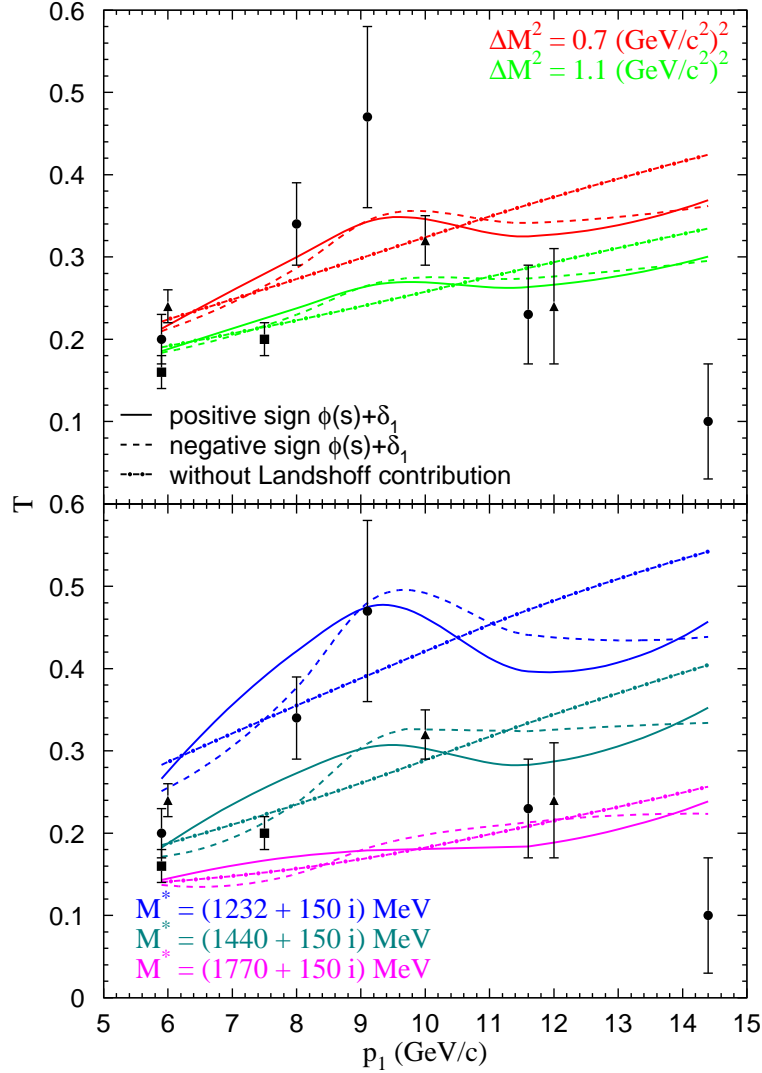}
\caption{The \nuc{12}{C}$(p,2p)$ transparency versus the incoming lab
  momentum $p_1$.  The upper (lower) panel depicts results using the
  FLFS (JM) model for CT.  Calculations including the effects of CT and
  NF with the positive (solid lines) and negative (dashed lines) sign
  for $\phi \left( s \right) + \delta_{1}$ are shown, along with the
  RMSGA+CT predictions (dot-dashed lines).  Data are from
  Refs.~\cite{carroll88} (solid triangles), \cite{mardor98} (solid
  squares), and \cite{leksanov01} (solid circles).}
\label{fig:trans.12c.FLFSvsJM}
\end{center}
\end{figure*}

\begin{figure*}[p]
\begin{center}
\includegraphics[width=0.75\textwidth]{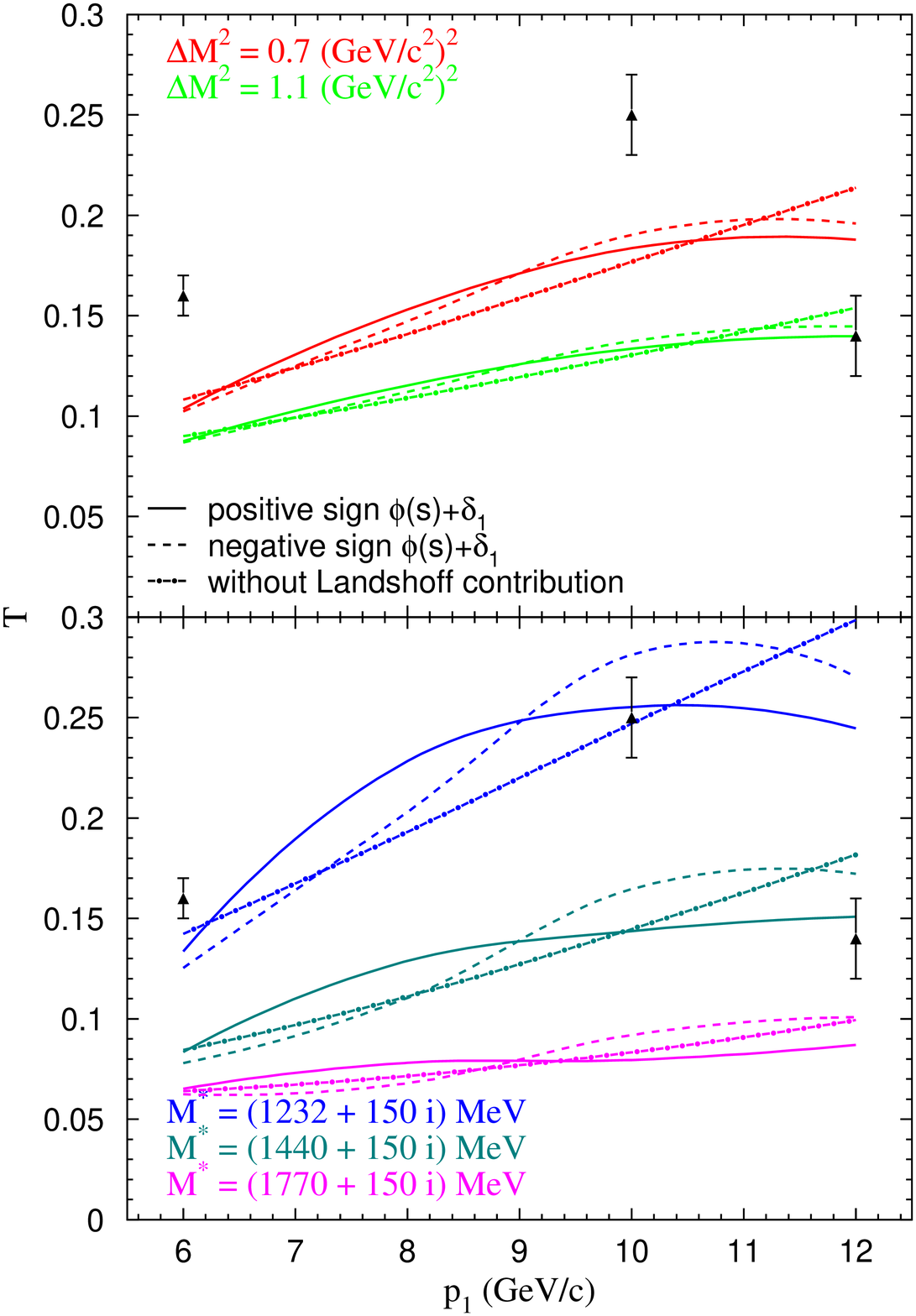}
\caption{As in Fig.~\ref{fig:trans.12c.FLFSvsJM}, but for \nuc{27}{Al}.
  Data are from Ref.~\cite{carroll88}.}
\label{fig:trans.27al.FLFSvsJM}
\end{center}
\end{figure*}

Figs.~\ref{fig:trans.12c.FLFSvsJM} and \ref{fig:trans.27al.FLFSvsJM} are
devoted to a comparison of the different CT models.  Results of the FLFS
quantum diffusion model are plotted for $\Delta M^2 = 0.7$ and
$1.1$~(GeV/c$^2$)$^2$.  For the $M^{*}$ parameter of the JM model we
consider three different values, representing the $\Delta$, the Roper
resonance, and the average of the lowest $P$-wave $N^{*}$ resonances.
For the imaginary part of $M^{*}$ a value of $150$~MeV was taken.  The
lowest values of the parameters $\Delta M^2$ and $M^{*}$ induce the
strongest increase of the RMSGA+CT transparency with the beam momentum
$p_1$ and also lead to the largest deviations between the predictions
including the NF mechanism and the corresponding RMSGA+CT results.

Fig.~\ref{fig:trans.12c.FLFSvsJM} shows that after including NF and CT,
the calculations correctly reproduce the maximum in the
\nuc{12}{C}$(p,2p)$ transparency at about $9.5$~GeV/c, but badly fail to
fall back low enough to account for the $14.4$~GeV/c data point.
Ralston and Pire \cite{ralston88}, on the other hand, do succeed in
reproducing the maxima and minima in the nuclear transparency data.
However, they assume that the transparency of the quark-counting term is
beam-energy independent, a rather speculative assumption.  The
CT-induced increase of the quark-counting transparency with energy
causes our RMSGA+CT+NF predictions to rise again at a momentum $p_{1}
\simeq 12$~GeV/c.

In the FLFS as well as the JM approach, the RMSGA+CT+NF predictions
reproduce the general trend of the data, but no variant achieves very
good agreement.  Furthermore, it is not possible to unambiguously
determine the value of the parameters $\Delta M^2$ and $M^{*}$, as the
best choice for these parameters depends on the target nucleus under
consideration.  Fig.~\ref{fig:trans.12c.FLFSvsJM} suggests that for the
\nuc{12}{C}$(p,2p)$ reaction $M^{*} = (1440 + 150 i)$~MeV leads to the
best agreement, while for the $\Delta M^2$ parameter no ``best'' choice
can be put forward.  Fig.~\ref{fig:trans.27al.FLFSvsJM}, on the other
hand, shows that for the \nuc{27}{Al} target nucleus the FLFS results
are systematically below the data in the region below $10$~GeV/c.  Using
$M^{*} = (1440 + 150 i)$~MeV in the JM model also does not increase the
transparency high enough so as to match the $6$ and $10$~GeV/c data
points, only with $M^{*} = (1232 + 150 i)$~MeV is the CT-induced
increase of the transparency strong enough.  

None of the results shown in this Letter include the color screening
effect (CSE) \cite{frankfurt85-88}.  This QCD effect suggests the
suppression of the small-size configurations in bound nucleons.  We deem
that the CT parameters are so badly constrained that controlling
additional mechanisms is out of reach for the moment.  In both the
RMSGA+CT and the RMSGA+CT+NF calculations, the inclusion of the CSE
decreases the transparency by $6$--$12\%$, with the largest effect
occuring at higher $p_1$.

Another effect that can be studied in Figs.~\ref{fig:trans.12c.FLFSvsJM}
and \ref{fig:trans.27al.FLFSvsJM} is the influence of the sign of $\phi
\left( s \right) + \delta_{1}$ on the RMSGA+CT+NF results.  For the FLFS
model of CT, the differences between calculations using both signs of
$\phi \left( s \right) + \delta_{1}$ are minor.  As already observed by 
Jennings and Miller \cite{jennings92b}, the results using the JM model
of CT are rather sensitive to this sign.  The discrepancy between the
FLFS- and JM-based calculations arises from the different structure of
the effective cross sections (\ref{eq:FLFS_eff_sigma}) and
(\ref{eq:JM_eff_sigma}).  Indeed, these effective cross sections not
only determine the attenuation of the quark-counting term in the nuclear
medium, but also the phase difference between the quark-counting and the
Landshoff term.  Whereas the real parts of both effective cross sections
are quite similar, the FLFS effective cross section
(\ref{eq:FLFS_eff_sigma}) is purely real, while its JM counterpart also
has an imaginary part.  This imaginary part causes the enhanced
sensitivity of the JM results to the sign of $\phi \left( s \right) +
\delta_{1}$.  As for which sign causes the best agreement with the data,
no firm conclusions can be drawn.  Indeed, the optimum choice for the
sign of $\phi \left( s \right) + \delta_{1}$ depends on the used CT
model (FLFS or JM), the value of the parameter $\Delta M^2$ or $M^{*}$,
and the target nucleus.  For the \nuc{7}{Li}$(p,2p)$ reaction, the
positive sign provides the best agreement, while the \nuc{63}{Cu}
transparency data rather require a negative sign.  

\begin{figure*}[p]
\begin{center}
\includegraphics[width=0.75\textwidth]{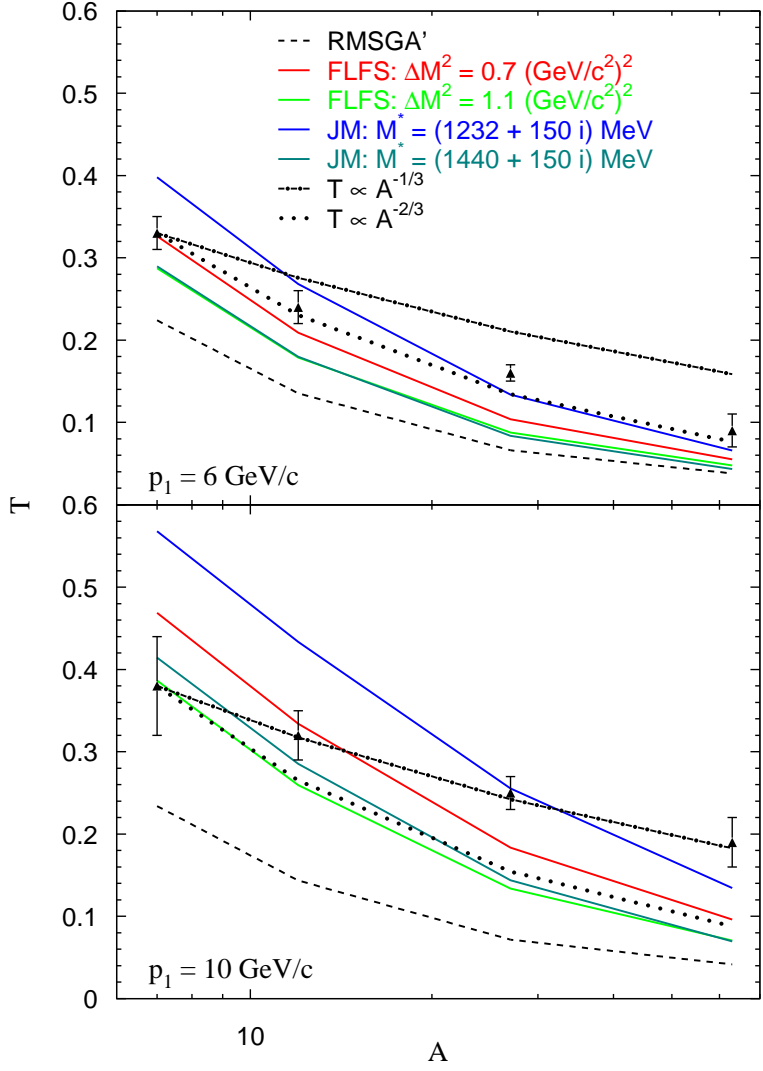}
\caption{$A$ dependence of the nuclear transparency at two values of the
  incoming lab momentum $p_1$.  The standard RMSGA calculations are
  represented by dashed curves, while the solid curves are RMSGA+CT+NF
  calculations with the positive sign for $\phi \left( s \right) +
  \delta_{1}$.  The solid curves correspond with different descriptions
  of the CT effects, as indicated by the legend.  The dot-dashed
  (dotted) curves display the $A^{-1/3}$ ($A^{-2/3}$) parametrization,
  normalized to the \nuc{7}{Li} data points.  Data are from
  Ref.~\cite{carroll88}.}
\label{fig:trans.Adep}
\end{center}
\end{figure*}

The $A$ dependence of the nuclear transparency at two values of the
incoming momentum $p_{1}$ is studied in Fig.~\ref{fig:trans.Adep}.  The
standard RMSGA calculations fall considerably below the data.  Further,
none of the RMSGA+CT+NF calculations succeed in simultaneously
describing the data for all target nuclei.  While the FLFS approach
agrees with the \nuc{7}{Li} and \nuc{12}{C} data points rather well
using $\Delta M^2 = 0.7~$(GeV/c$^2$)$^2$, the FLFS results tend to
underestimate the \nuc{27}{Al} and \nuc{63}{Cu} data.  With regard to
the $M^{*}$ parameter of the JM model, for the \nuc{7}{Li} and
\nuc{12}{C} nuclei a value of $M^{*} = (1440 + 150 i)$~MeV seems
acceptable, whereas the heavier \nuc{27}{Al} and \nuc{63}{Cu} nuclei
need a smaller $M^{*}$ value.  A general feature of the RMSGA+CT+NF
predictions is that their $A$ dependence is steeper than the data.
Finally, the dot-dashed and dotted curves indicate that the $6$~GeV/c
data are proportional to $A^{-2/3}$, while at $10$~GeV/c the $A$
dependence of the data is more gradual ($T \propto A^{-1/3}$).  This
trend is not reproduced by the standard RMSGA predictions, which are
almost independent of the incoming momentum.  When CT and NF effects are
included, the softening of the $A$ dependence with increasing incoming
momentum is also present in the calculations.  For example, the FLFS
calculations with $\Delta M^2 = 0.7~$(GeV/c$^2$)$^2$ correspond to $T
\propto A^{-0.83}$ at $6$~GeV/c incoming momentum and $T \propto
A^{-0.70}$ at $10$~GeV/c.

\section{Conclusions}
\label{sec:concl}

In conclusion, we have developed a relativistic framework to calculate
the nuclear transparency for $A(p,2p)$ processes.  A relativistic
multiple-scattering Glauber model was used to account for the IFSI.  To
reduce the computational cost of the RMSGA calculations, some additional
approximations were made.  The predictions with the full and the
approximated RMSGA approach agree at the few percent level.  Thus, to
determine integrated quantities such as the nuclear transparency, a
valid alternative for the computationally intensive RMSGA framework is
available.

Using the concept of nuclear filtering, our calculations are in
qualitative agreement with the data, thereby confirming earlier
calculations \cite{jennings92b,jennings92a,jennings93}.  Furthermore,
our calculations seem to indicate that CT is imperative to increase the
calculations to the level of the data.  The same conclusion was reported
in Refs.~\cite{jennings93,jain93}.  The quantitative description of the
data, however, is far from perfect and it is not possible to constrain
the magnitude of the parameters in the CT models.  

The indications for CT in $A(p,2p)$ reactions are not necessarily in
contradiction with the results from $A(e,e'p)$ experiments.  Although
the $A(e,e'p)$ transparencies show no significant increase with the
four-momentum transfer $Q^2$ \cite{eep_exp} and can be reasonably
reproduced in the RMSGA framework, the existence of CT can not be
excluded since the predicted effect is small \cite{eep_theory}.  This is
caused by the small expansion times of the PLC to normal size at the
present $A(e,e'p)$ kinematics.  The effect of CT is more pronounced in
the $A(p,2p)$ transparency for different reasons.  First, in the
$A(p,2p)$ reaction there are three particles that can experience CT
instead of only one in $A(e,e'p)$ reactions.  Second, $A(p,2p)$ data are
available up to $Q^2 = |t|$ values of $12.7$~(GeV/c)$^2$, while
$A(e,e'p)$ transparency experiments are restricted to $Q^2 \lesssim
8$~(GeV/c)$^2$.

A number of uncertainties involving the $A(p,2p)$ transparency remain.
One subject of discussion is the size of the Landshoff term.  According
to Botts et al. \cite{bottsQCD}, this term might also be small-sized.
This would make the survival probability of the Landshoff and the
quark-counting term rather similar and would weaken the oscillations in
the energy dependence of the computed transparencies.  Apart from the
Ralston-Pire picture discussed above, other explanations of the energy
dependence of the transparency have been suggested.  For example,
Brodsky and de Teramond \cite{brodsky88} interpreted the oscillatory
behavior in terms of two broad baryon resonances associated with strange
and charmed particle production thresholds, interfering with a
perturbative QCD background.  An improved set of data, particularly at
higher energies, is essential to clarify these issues.  The $50$~GeV
proton synchrotron that is under construction at J-PARC \cite{j-parc}
opens great opportunities for this research.

\begin{ack}
The authors would like to thank M. Sargsian for providing us with the
parametrization of the free $pp$ elastic cross section.  This work was
supported by the Fund for Scientific Research, Flanders (FWO).
\end{ack}



\end{document}